\documentclass[a4paper]{article}

\usepackage{spconf}

\usepackage{tikz}

\usepackage{multirow}
\usepackage{subcaption}

\usepackage{graphicx}
\usepackage{amssymb,amsmath,bm}
\usepackage{textcomp}

\ninept

\title{Personalized Acoustic Modeling by Weakly Supervised Multi-task Deep Learning Using Acoustic Tokens Discovered from Unlabeled Data}

\makeatletter
\def\name#1{\gdef\@name{#1\\}}
\makeatother \name{{\em Cheng-Kuan Wei$^1$, Cheng-Tao Chung$^1$, Hung-Yi Lee$^2$ and Lin-Shan Lee$^2$}}

\address{$^1$Graduate Institute of Electrical Engineering, National Taiwan University \\
  $^2$Graduate Institute of Communication Engineering, National Taiwan University \\
  {\small \tt r02921036@ntu.edu.tw, f01921031@ntu.edu.tw, tlkagkb93901106@gmail.com, lslee@gate.sinica.edu.tw}
  \thanks{Copyright 2017 IEEE. Published in the IEEE 2017 International Conference on Acoustics, Speech, and Signal Processing (ICASSP 2017), scheduled for 5-9 March 2017 in New Orleans, Louisiana, USA. Personal use of this material is permitted. However, permission to reprint/republish this material for advertising or promotional purposes or for creating new collective works for resale or redistribution to servers or lists, or to reuse any copyrighted component of this work in other works, must be obtained from the IEEE. Contact: Manager, Copyrights and Permissions / IEEE Service Center / 445 Hoes Lane / P.O. Box 1331 / Piscataway, NJ 08855-1331, USA. Telephone: + Intl. 908-562-3966.}
}

\begin{document}

  \fontsize{9.}{9.3}\selectfont  

  \maketitle
  
  \begin{abstract}
    It is well known that recognizers personalized to each user are much more effective than user-independent recognizers. With the popularity of smartphones today, although it is not difficult to collect a large set of audio data for each user, it is difficult to transcribe it. However, it is now possible to automatically discover acoustic tokens from unlabeled personal data in an unsupervised way. We therefore propose a multi-task deep learning framework called a phoneme-token deep neural network (PTDNN), jointly trained from unsupervised acoustic tokens discovered from unlabeled data and very limited transcribed data for personalized acoustic modeling. We term this scenario ``weakly supervised''. The underlying intuition is that the high degree of similarity between the HMM states of acoustic token models and phoneme models may help them learn from each other in this multi-task learning framework. Initial experiments performed over a personalized audio data set recorded from Facebook posts demonstrated that very good improvements can be achieved in both frame accuracy and word accuracy over popularly-considered baselines such as fDLR, speaker code and lightly supervised adaptation. This approach complements existing speaker adaptation approaches and can be used jointly with such techniques to yield improved results.
  \end{abstract}
  \noindent{\bf Index Terms}: speech adaptation, unsupervised learning, deep neural network, multitask learning, transfer learning

  \section{Introduction}
    Today most commercially available speech recognizers are user-independent, although it is well known that recognizers personalized to each individual user offer superior performance, because the speaker characteristics and language patterns of each individual user are captured in the recognition models. With the popularity of smartphones today, collecting personal audio data for each individual user is not difficult; annotating this data, however, is difficult. If for each individual user we could use a commercially-available user-independent recognizer to obtain a small quantity of his or her personal audio data (e.g., 10 to 50 utterances) and then make the necessary corrections to this data, this small bit of annotated data could be used together with his or her other much larger set of unlabeled personal audio data to train personalized acoustic models. This ``weakly supervised''scenario is the focus of this paper.

    Speaker adaptation has been investigated thoroughly not only in the past, but also recently, in particular within the years after the deep learning paradigm became mainstream in the global speech technology community. Improved regularization approaches have been developed for the adaptation training criterion \cite{stadermann2005two}\cite{li2006regularized}\cite{yu2013kl}. Supplementary features are appended to the input to compensate for different acoustic conditions, as with i-vectors \cite{senior2014improving}\cite{saon2013speaker}, underlying factors in joint factor analysis (JFA) \cite{li2014factorized}, or the sequence summarizing neural network (SSNN) \cite{vesel2016sequence}. DNN's are also trained with a set of automatically-obtained speaker-specific features referred to as speaker codes \cite{abdel2013fast}\cite{xue2014direct}. Meanwhile, many groups use transformation-based schemes that treat speaker-independent (SI) neural networks as canonical models while adding additional linear hidden layers as speaker-dependent (SD) transformations either prior to the input layer -- sometimes referred to as feature-discriminative linear regression (fDLR) \cite{seide2011feature} -- or prior to the hidden layer \cite{gemello2007linear}\cite{yao2012adaptation} or to the output layer \cite{li2010comparison}. Alternatively, instead of modeling such additional transformations, the Hermitian-based activation function is used in adaptation while keeping the DNN weights fixed \cite{siniscalchi2012hermitian}. However, these methods all rely primarily on annotated adaptation data and do not take into account the abundant quantities of unlabeled personalized data.

    Conventional approaches to use unlabeled data include unsupervised or ``lightly supervised'' adaptation, very often considered in low-resource speech recognition \cite{lamel2002unsupervised}\cite{lamel2002lightly}\cite{wessel2005unsupervised}\cite{ma2008unsupervised}. The basic idea is to use a speaker-independent (SI) model or a background model to transcribe the unlabeled raw audio data and generate approximate transcriptions used for training, sometimes as an iterative process. Further improvements are possible by for instance removing utterances with less reliable transcriptions or by selecting utterances with more reliable transcriptions \cite{kapralova2014big}\cite{doulaty2015data}, either based on confidence scores and other useful features, or by using models such as conditional random fields (CRF) \cite{sheng2015discriminative}\cite{li2016data}. In these approaches, all knowledge which can be extracted from the unlabeled data is based on either the SI or background model, or the limited annotated data. This raises the question: is there any knowledge that can be extracted directly from the unlabeled data?

    In recent years it has become clear that acoustic tokens can be automatically discovered from unlabeled corpora in an unsupervised fashion \cite{chung2013unsupervised}\cite{chung2014unsupervised}\cite{park2008unsupervised}\cite{jansen2011towards}; these tokens have been shown successful for both spoken term detection and spoken document retrieval tasks \cite{li2013towards}\cite{lee2013enhancing}\cite{jansen2010towards}. This implies that these automatically discovered acoustic tokens correlate well to underlying linguistic units such as phonemes.
    
    The acoustic models for these tokens can be trained from unlabeled data, which in turn can be decoded into sequences of these tokens. Thus these tokens are extracted directly from the unlabeled data without using any other knowledge. It is therefore reasonable to consider if such tokens can be used jointly with the limited annotated data in the weakly supervised scenario considered here.

    One major problem is that the direct relationship between these automatically discovered acoustic tokens and the phoneme labels is unknown and likely to be noisy. This problem may be solved by multi-task learning, in which multiple related tasks are trained simultaneously, for example with shared hidden layers in DNN, and thus benefit from each another. Multi-task learning has been shown to offer significant improvements in multilingual acoustic models because of cross-language knowledge transfer \cite{huang2013cross}\cite{heigold2013multilingual}. Such knowledge transfer across tasks is helpful for many reasons: local optima supported by more tasks may be better, more data can be used to learn the same set of parameters, some knowledge may be easier to learn in one task than in others, and so on \cite{thrun1996learning}\cite{kovac2005multitask}.

    In this paper, we propose a multi-task deep learning framework for the weakly supervised personalized acoustic modeling scenario mentioned above. In this framework, unlabeled data are transcribed into sequences of automatically discovered tokens, and this knowledge obtained straight from the unlabeled data is used jointly with the phonemes in the annotated data for multi-task learning, or by a DNN with shared hidden layers. The underlying basic idea is that the high degree of similarity between the HMM states of the unsupervised token models and the supervised phoneme models may be mutually beneficial during multi-task learning. If we consider the unlabeled data with its corresponding token sequences to be a different language, then we can view the task as cross-language knowledge transfer for multi-lingual acoustic modeling. In Section 2 we present the proposed approach, and in Sections 3 and 4 we go through the experiments
    and results.

  \section{Proposed Approach}

  \subsection{Automatic Discovery of Acoustic Tokens from Unlabeled Data}
    \label{subsec:acoustic_token}

    Acoustic tokens, which refer to short segments of sounds frequently occurring in a corpus, are discovered automatically by machine and are very similar to human-defined phonemes. Here we seek to discover automatically such acoustic tokens \cite{chung2013unsupervised}\cite{chung2014unsupervised} from the personal audio data of each individual user. This can be achieved in the following way. Let $\overline{O}$ represent the acoustic feature vector sequences for the entire corpus (the audio data of a user). We begin with an initial set of tokens and the initial token label sequence $W_0$, including boundaries for each token for the observation $\overline{O}$ as in (\ref{eq:token_train_eq1}) below. We first use segmentation algorithms to divide the utterances in $\overline{O}$ into small signal segments based on discontinuities in the contours of energy- and MFCC-related features. We then compute the mean of the MFCC vectors for each of these small signal segments, and perform K-means clustering for the mean vectors over the whole corpus $\overline{O}$. We then assign to each cluster a token ID (this is the initial token set), with which we define $W_0$, the initial token label segments, including the boundaries over $\overline{O}$. We then fine-tune this initial label $W_0$ using the following iterative optimization approach: In each iteration $i$, given the label $W_{i-1}$ over $\overline{O}$, we train the HMM model for each token using the short signal segments with the given token ID using the Baum-Welch algorithm. This yields a set of HMMs for all the tokens with parameters $\theta_i$ as in (\ref{eq:token_train_eq2}), which we then use to decode the whole corpus $\overline{O}$ to obtain a new label $W_i$ as in (\ref{eq:token_train_eq3}). We then repeat (\ref{eq:token_train_eq2})(\ref{eq:token_train_eq3}) iteratively until the
    convergence of the generated labels $W_i$, including the label boundaries.

    \begin{equation}
    W_0 = \mathit{initialization}(\overline{O})
    \label{eq:token_train_eq1}
    \end{equation}

    \begin{equation}
    \theta_i = \arg\max_\theta P(\overline{O}|\theta, W_{i-1})
    \label{eq:token_train_eq2}
    \end{equation}

    \begin{equation}
    W_i = \arg\max_W P(\overline{O}|\theta_i, W)
    \label{eq:token_train_eq3}
    \end{equation}

    There are two key parameters for the token HMMs. The
    number of states in each token HMM, $m$, controls the token lengths, 
    or the temporal granularity of the tokens. The initial total number of the clusters mentioned above or the distinct number of tokens, $n$, concerns the segmentation of the phonetic space, or the phonetic granularity of the tokens. Hence each parameter set $\psi = (m, n)$ defines a set of tokens learned from $\overline{O}$.

    It is difficult to know the ideal parameter set for a given corpus, although when examining our data we noted the token set for $\psi = (3, 100)$ approximated gender-dependent phonemes and the token set for $\psi = (13, 300)$ roughly approximated syllables. Thus we more or less capture the characteristics and behavior of the underlying language described by the corpus by using multi-granular acoustic tokens, that is, by combining acoustic tokens with a variety of model granularities.

  \subsection{Phoneme-Token DNN (PTDNN)}
      \begingroup
    \makeatletter
    \renewcommand{\p@subfigure}{}
    \makeatother
    \begin{figure}[!ht]

        \begin{subfigure}[t]{0.45\linewidth}
            \centering
            \caption{}
            \includegraphics[width=\linewidth]{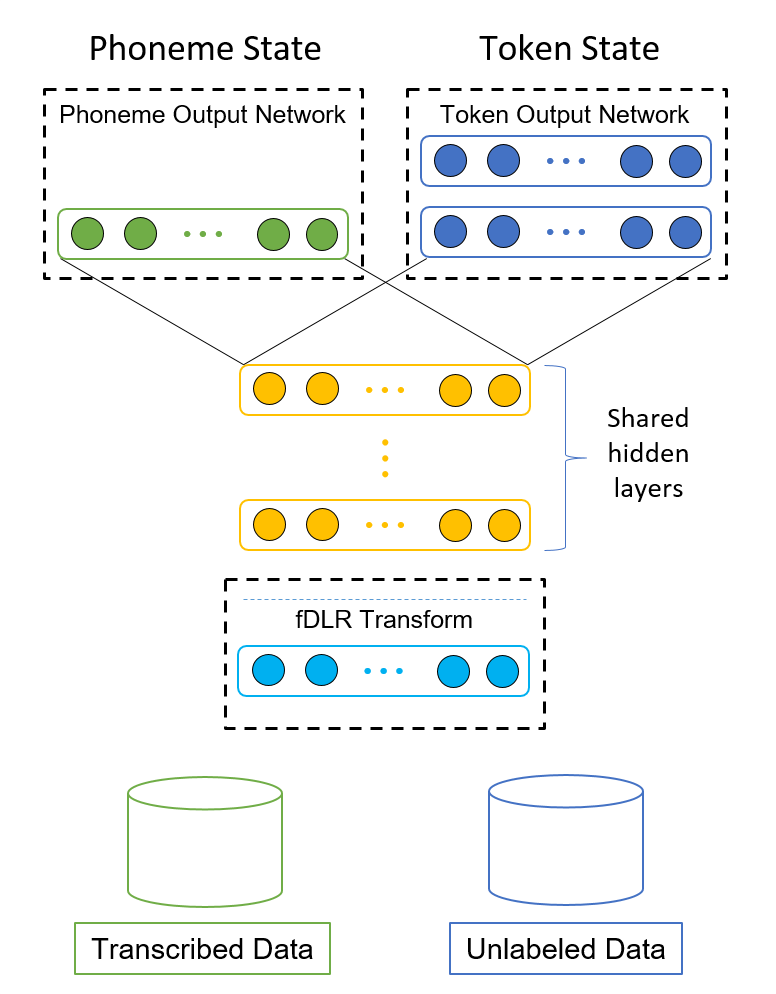}
            \label{fig:architecture}
        \end{subfigure}
        \quad
        \begin{subfigure}[t]{0.45\linewidth}
            \centering
            \caption{}
            \includegraphics[width=\linewidth]{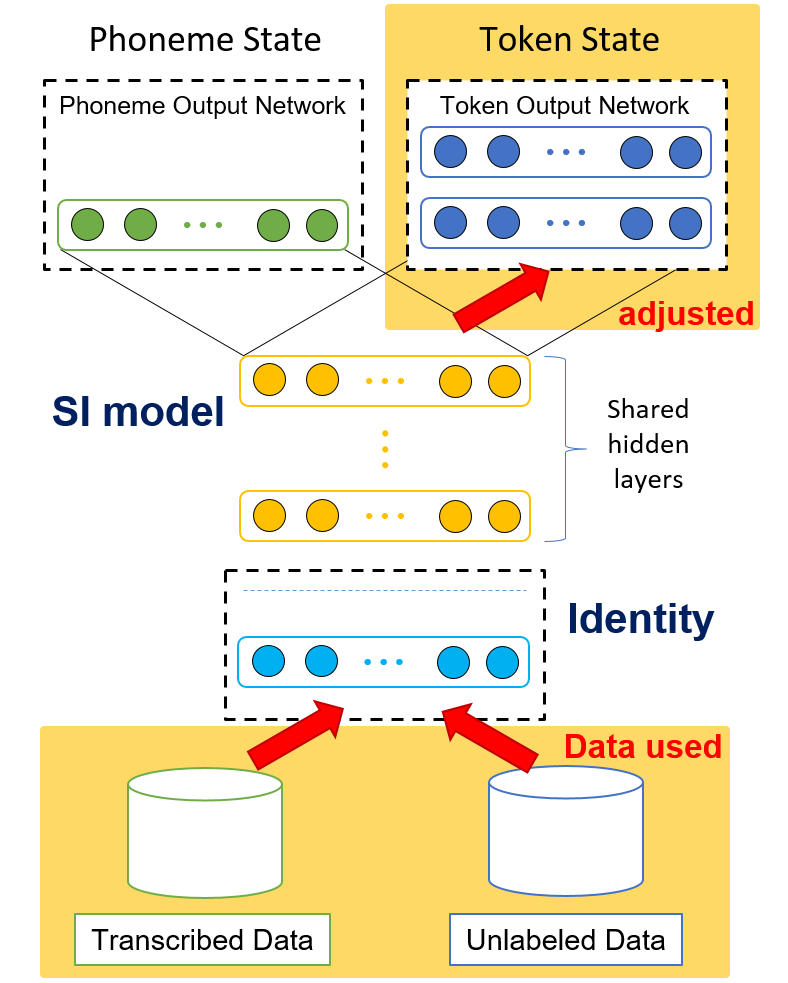}
            \label{fig:init}
        \end{subfigure}

        \centering
        \begin{subfigure}[t]{0.45\linewidth}
            \centering
            \caption{}
            \includegraphics[width=\linewidth]{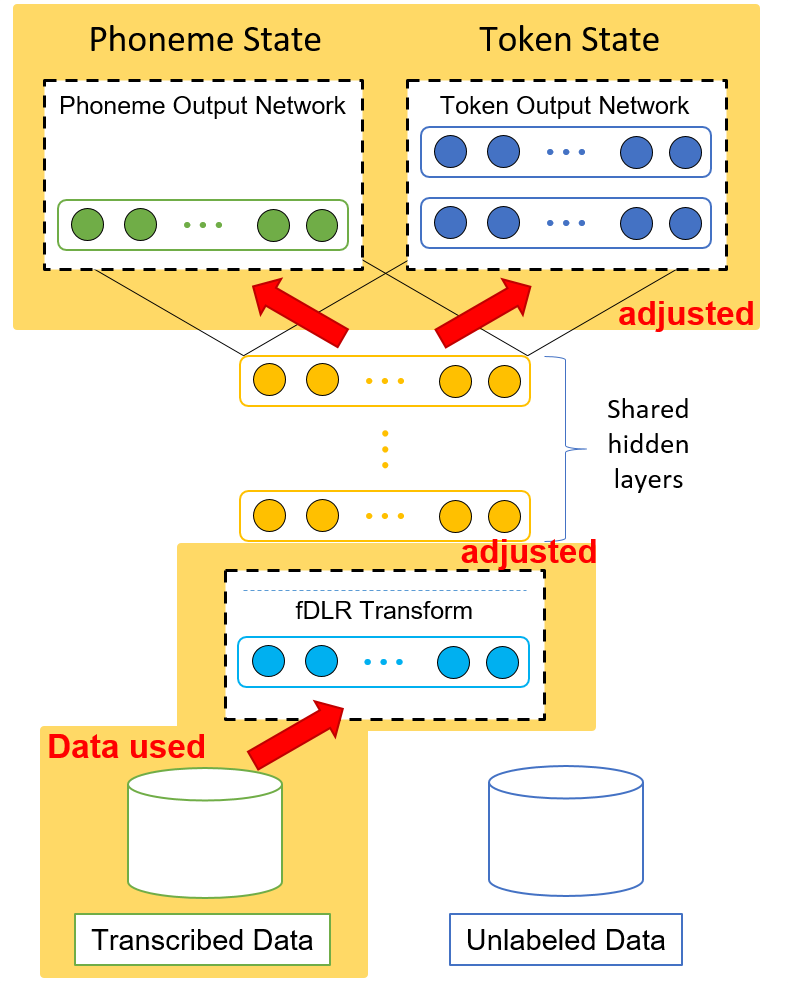}
            \label{fig:joint_opt1}
        \end{subfigure}
        \quad
        \begin{subfigure}[t]{0.45\linewidth}
            \centering
            \caption{}
            \includegraphics[width=\linewidth]{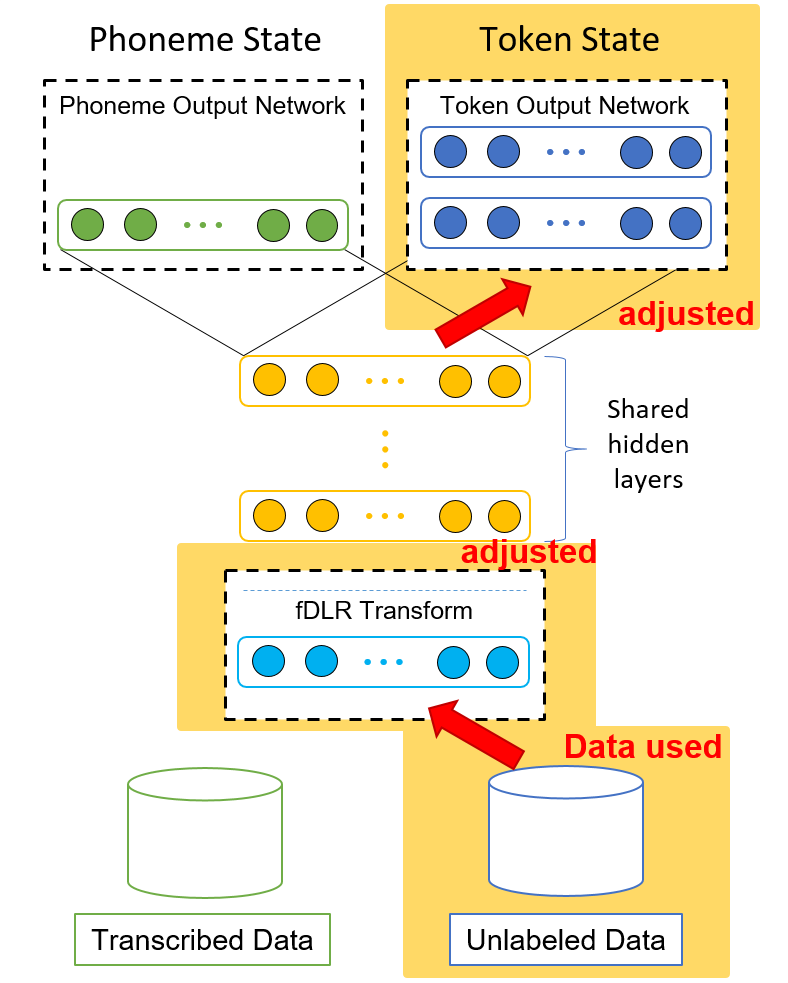}
            \label{fig:joint_opt2}
        \end{subfigure}
        \caption{(\ref{fig:architecture}) The phoneme-token DNN (PTDNN); (\ref{fig:init}) Training step 1: initialization; (\ref{fig:joint_opt1})(\ref{fig:joint_opt2}) Training step 2: joint optimization.}
        \label{fig:pt_dnn}

    \end{figure}
    \endgroup

    The phoneme-token DNN (PTDNN) proposed in this paper is depicted in Fig.~\ref{fig:pt_dnn}(\ref{fig:architecture}). At the top of this figure, the states for each phoneme HMM (in the left) and each token HMM (on the right) are the two tasks to be learned in parallel with a set of shared hidden layers and a shared feature discriminative linear regression (fDLR) transformation. The phoneme state probabilities are learned from the transcribed data (bottom left), while the token state probabilities are learned from the unlabeled data (bottom right) along with the transcribed data (bottom left). If, as suggested above, the acoustic tokens are considered a different language, this corresponds to the structure for multilingual acoustic modeling. Furthermore, acoustic token sets of multiple granularities $\psi = (m, n)$, or other features or labels, can also be learned jointly in this architecture, simply by adding more targets and the corresponding output networks in Fig.~\ref{fig:pt_dnn}(\ref{fig:architecture}).

  \subsection{Speaker Adaptation}
    \label{subsec:speaker_adaptation}
    Speaker adaptation involves the three steps summarized below and is shown in Fig.~\ref{fig:pt_dnn}(\ref{fig:init})(\ref{fig:joint_opt1})(\ref{fig:joint_opt2}).

    \begin{enumerate}
        \item Initialization

        As shown in Fig.~\ref{fig:pt_dnn}(\ref{fig:init}), we first take a speaker-independent (SI) DNN-HMM acoustic model trained on the large set of speaker-independent data, and use its hidden layers for the shared hidden layer needed for initialization. We also initialize the fDLR transformation with an identity matrix. With these parameters in the shared hidden layers and the fDLR transformation all fixed in this step, we train the token output network only (upper right corner) on both the transcribed and unlabeled data of the personalized audio data set, where the transcribed data are also decoded into tokens to be used as the token state target.

        \item Joint optimization

        After initialization, we then jointly optimize the phoneme state and the token state output network iteratively. As in Fig.~\ref{fig:pt_dnn}(\ref{fig:joint_opt1}), we first use the limited set of transcribed data (bottom left) to train both the phoneme and the token targets with the objective function of (\ref{eq:joint_opt_eq1}) being the weighted sum of that for the two targets, and then use the large set of unlabeled data to train the acoustic token state only with an objective function of (\ref{eq:joint_opt_eq2}), as depicted in Fig.~\ref{fig:pt_dnn}(\ref{fig:joint_opt2}). In this way, we attempt to optimize the output networks of both targets synchronously and jointly. Thus we shuffle the transcribed and unlabeled data at a mini-batch level to facilitate the mutual learning of the included knowledge.

        \begin{equation}
        f = W_{\mathit{phoneme}} \cdot f_{\mathit{phoneme}}+W_{\mathit{token}} \cdot f_{\mathit{token}}
        \label{eq:joint_opt_eq1}
        \end{equation}

        \begin{equation}
        f = W_{\mathit{token}} \cdot f_{\mathit{token}}
        \label{eq:joint_opt_eq2}
        \end{equation}

        \item Transferring back

        Finally, in the last step (not shown in the figure), in order to emphasize the desired phoneme state target, we further optimize the phoneme state output network only to transfer all the knowledge learned back, for phoneme recognition and to fine-tune the model.

    \end{enumerate}

    In the whole training procedure, to prevent overfitting on the very small training set, we update only the parameters of the fDLR transformation and the output networks for phoneme and token states. As the size of the training set increases, we can then attempt to adjust more parameters, or even omit the above initialization step and just begin with step two to better fit the data.

  \section{Experimental Setup}

    To better simulate the personalized recognizer scenario mentioned in the introduction, the experiments were performed on a Facebook post corpus we collected. Each of five male and five female speakers was asked to produce 1000 utterances, all extracted from his or her own Facebook posts in a spontaneous speech style; this yielded a 6.6-hour dataset. These utterances were primarily in Chinese but about 4.1\% of the words were in English. We divided the 1000 utterances for each speaker into three sets: 500 utterances as the adaptation set, 250 as the development set, and 250 for testing. Also, we randomly selected as the transcribed data 50 utterances out of the 500 in the adaptation set.

    The initial speaker-independent (SI) model was trained using a mixed corpus of the ASTMIC corpus (read speech in Mandarin, 31.8 hours) \cite{lee2014spoken} and the EATMIC corpus (read speech in English produced by Taiwanese speakers, 29.7 hours) \cite{lee2014spoken} with 4 hidden layers, each with 2048 units. The acoustic features used were the 13-dimensional MFCCs plus their first and second order delta features. The features were normalized to zero mean and unit variance, and a context window of 9 frames (4 frames on each side) was used. A trigram language model was used in the decoding, which was trained on data crawled from the PTT bulletin board system (BBS), a popular system in Taiwan with more than 1.5 million registered users and over 20000 new posts daily. Before training the model, we generated the personal acoustic token sets for each speaker using all 500 utterances of the adaptation data for the speaker. For the parameter set $\psi = (m, n)$, where $m$ is the number of states in each token HMM and $n$ is the number of distinct tokens in the set, we set $m=3, 5, 9, 13$ and $n=50, 100, 200, 300$, aiming for 16 sets of acoustic tokens, each with a different granularity.

    As in Section~\ref{subsec:speaker_adaptation} and Fig.~\ref{fig:pt_dnn}(\ref{fig:init})(\ref{fig:joint_opt1})(\ref{fig:joint_opt2}), we initialized the token output network using the state-aligned labels for the given acoustic token set. We pretrained the token output network with a stacked RBM, and then fine-tuned the output network for 100 epochs with a relatively large learning rate of 0.01, after which we jointly optimized the output networks for phoneme and token states for 50 epochs iteratively with the learning rate set to $10^{-3}$, $W_{phoneme}=4$, and $W_{token}=1$. Finally, we transferred the knowledge learned back to the phoneme output network for 50 epochs with a learning rate of $10^{-4}$.

  \section{Experimental Results}

  \subsection{Speaker Adaptation Experiment}

    \begin{table}[!h]
    \centering
    \resizebox{\columnwidth}{!}{
    \begin{tabular}{|c|l|c|c|}
    \hline
    \multicolumn{2}{|c|}{Models}                        & Frame accuracy   & Word accuracy    \\ \hline
    (A)   & SI (DNN-HMM)                                & 31.91\%          & 57.45\%          \\ \hline
    (B-1) & fDLR                                        & 41.66\%          & 65.70\%          \\
    (B-2) & Speaker code                                & 42.11\%          & 65.92\%          \\ \hline
    (C)   & Lightly supervised adaptation               & 45.04\%          & 62.10\%          \\ \hline
    (D-1) & PTDNN, $\psi=(5,200)$                       & 43.49\%          & 66.94\%          \\
    (D-2) & [PTDNN, $\psi=(5,200)$] + fDLR              & \textbf{48.74}\% & \textbf{69.83\%} \\ \hline
    \end{tabular}
    }
    \caption{Basic speaker adaptation results for the proposed PTDNN and the baselines}
    \label{tab:vs_baseline}
    \end{table}

    As mentioned above, we used all the 500 utterances of adaptation data for each speaker to generate the personal token sets. We also randomly selected a subset of 50 utterances out of the 500 as the transcribed data, and used the other 450 as the unlabeled data. The test was performed on the test set of 250 utterances, disjoint from the adaptation set for each speaker.

    The results are listed in Table~\ref{tab:vs_baseline}. As baselines we also include the SI model with DNN-HMM (row A), fDLR speaker adaptation (row (B-1)), speaker code adaptation (row (B-2)), and lightly supervised adaptation \cite{lamel2002unsupervised}\cite{lamel2002lightly} (row (C)). In fDLR and speaker code in rows (B-1) and (B-2), only the 50 utterances of transcribed data were used in adaptation, while the remaining 450 utterances of unlabeled data were not used. In the lightly supervised adaptation in row (C), the 450 unlabeled utterances were first recognized by the SI models to produce machine-generated transcriptions. These data were then combined with the 50 utterances of transcribed data with human-generated transcriptions to form a complete set of 500 utterances for adaptation. We see that both fDLR and the speaker code improved significantly both the frame and word accuracies over the SI model (row (B-1)(B-2) vs (A)), while speaker code was slightly better than fDLR in the scenario considered here (rows (B-2) vs (B-1)). We also see that lightly supervised adaptation yielded similar improvements over SI models (rows (C) vs (A)), although it slightly underperformed fDLR and speaker code in word accuracy (rows (C) vs (B-1)(B-2)) but was better in frame accuracy (rows (C) vs (B-1)(B-2)), in the scenario considered, probably due to the added 450 utterances of adaptation data (which yielded better frame accuracy) and the errors in the machine-generated transcriptions (incorrect phoneme-to-word mappings can degrade word accuracy).

    In row (D-1) is the proposed approach (PTDNN) with the parameter set $\psi=(5,200)$ using 50 utterances of transcribed data and 450 utterances of unlabeled data. This parameter set was chosen from all the sets tested and yielded the best results. It is likely that this set of tokens $(5,200)$ better modeled the linguistic units in the corpus tested (phonemes or high-frequency syllables, for example). We see that PTDNN outperformed fDLR (rows (D-1) vs (B-1)) by 1.83\% in frame accuracy and 1.24\% in word accuracy absolutely. It was also much better than speaker code in a similar way (rows (D-1) vs (B-2)), except by a slightly smaller range. Note that the proposed approach in row (D-1) did not suffer from the coarse tokens discovered from the unlabeled data, because the fine phoneme states and the coarse token states were jointly learned and converged in the iterative training step. PTDNN also outperformed lightly supervised adaptation (rows (D-1) vs (C)) in word accuracy (by 4.84\% absolutely), although it was slightly worse in frame accuracy, probably because the 450 utterances of unlabeled data were transcribed by a supervised SI model in row (C) but used to produce the token set only in an unsupervised way in row (D-1). Thus the frame-level phoneme information that was used was more precise.

    We further integrated PTDNN with $\psi=(5,200)$ in row (D-1) with fDLR in row (B-1) via a weighted summing of the output state posteriors (weights selected with the development set) with the results in row (D-2). This resulted in the best performance in this series of experiments: it improved the frame accuracy by 7.08\% and word accuracy by 4.13\% over taking fDLR alone as the baseline (rows (D-2) vs (B-1)). This verified that PTDNN is complementary to other adaptation approaches and capable of yielding additive improvements.

  \subsection{Granularity parameter sets $\psi=(m,n)$}

    \begin{figure}[!ht]
      \centering
      \includegraphics[width=\linewidth]{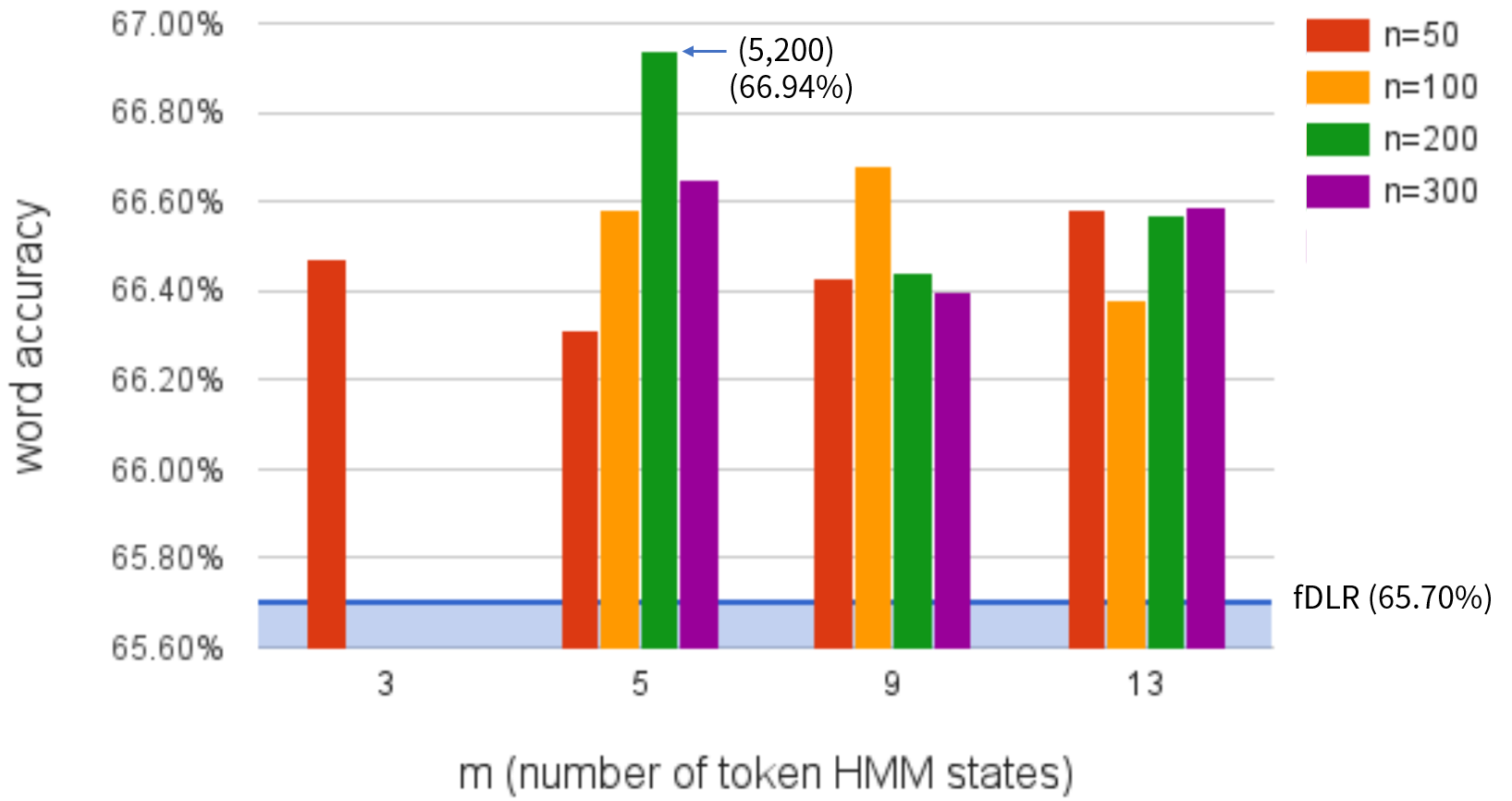}
      \caption{Word accuracies for the proposed PTDNN with tokens sets for different parameters $\psi=(m,n)$}
      \label{fig:diff_granularity}
    \end{figure}

    As mentioned above, there are many choices for the granularity parameter set $\psi=(m,n)$. In the experiments we chose $m=3, 5, 9, 13$ and $n=50, 100, 200, 300$. Out of the resulting 16 combinations of $\psi=(m,n)$, the token sets with $\psi=(3,100), (3,200), $ and $(3,300)$ would not converge when generating the tokens in Equations~(\ref{eq:token_train_eq2}) and (\ref{eq:token_train_eq3}) of Section~\ref{subsec:acoustic_token}, probably because such short models with $m=3$ was too short for phonemes, but for a single speaker $n=100, 200,$ or $300$ exceeded the number of phonemes by too much. As a result, many different redundant tokens were chosen, and therefore the sets did not converge.

    The remaining 13 sets of parameters were successfully used in the PTDNN training. The word accuracies obtained with 50 utterances of transcribed data and 450 unlabeled data, corresponding to the last column of Table~\ref{tab:vs_baseline}, are shown in Fig.~\ref{fig:diff_granularity}. The blue horizontal line on the bottom is for the fDLR baseline (65.70\%, row (B-1) of the table), while the highest bar is for $\psi=(5,200)$ (66.94\%, row (D-1)).

    We observe that all the token sets offered reasonable improvements over fDLR regardless of the choice of $(m,n)$; this suggests that the proposed approach is robust to the choice of these parameters. However different token sets yielded slightly different performance. The average performance of the 13 sets of tokens was 66.54\%, 0.84\% higher than fDLR absolutely. These results verify that these multi-granular tokens did indeed capture differing speaker-specific phonetic characteristics or information from the unlabeled data, and therefore that the approach was helpful here.

    Also, we note that the better token sets may have to do with the phonetic structures of the underlying language for the data. For example, for the best token set $(5, 200)$, $m=5$ could be a good number to model tokens close to phonemes, while $n=200$ was not too far from the order of Chinese phonemes plus English phonemes. On the other hand, $m=13$ could be a good number to model syllables. The majority of the data were in Mandarin, a syllable-based language. Of the roughly 400 Mandarin syllables, about 200 are frequently used. This may be why in Fig.~\ref{fig:diff_granularity} with $m=13$, the results were consistent and reasonably good for most cases.

  \subsection{More or Less Transcribed Data and More Token Sets}
  
    \begin{table}[!h]
    \centering
    \resizebox{\columnwidth}{!}{
    \begin{tabular}{|c|l|c|c|c|}
    \hline
    \multicolumn{2}{|c|}{Word accuracy}            & \multicolumn{3}{|c|}{Number of transcribed utterances}   \\ \hline
    \multicolumn{2}{|c|}{Models}                   & 10      & 50      & 100                                  \\ \hline
    (A)   & SI (DNN-HMM)                           & \multicolumn{3}{|c|}{57.45\%}                            \\ \hline
    (B-1) & fDLR                                   & 60.08\% & 65.70\% & 67.73\%                              \\
    (B-2) & Speaker code                           & 60.34\% & 65.92\% & 67.89\%                              \\ \hline
    (C)   & Lightly supervised adaptation          & 61.45\% & 62.10\% & 63.18\%                              \\ \hline
    (D-1) & Proposed, $m=5$, $n=200$               & 63.05\% & 66.94\% & 68.65\%                              \\ 
    (D-2) & Proposed with 4 token sets             & \textbf{64.89}\% & \textbf{67.15}\% & \textbf{68.75}\%   \\ \hline
    \end{tabular}
    }
    \caption{Word accuracy for more or less transcribed data and more token sets}
    \label{tab:diff_utt}
    \end{table}

    We are also curious to know what happens when given greater or fewer numbers of transcribed utterances. In addition to the 50 utterances of transcribed data from Table~\ref{tab:vs_baseline}, in Table~\ref{tab:diff_utt} we list more word accuracy results when given 10 and 100 utterances of transcribed data (and 490 and 400 of unlabeled data respectively, always making 500 utterances of personalized data in total). The middle 50-utterance column results are copied over from Table~\ref{tab:vs_baseline}, and two columns of results for 10 and 100 utterances are added on either side. Here rows (A)(B-1)(B-2)(C)(D) are copied from rows (A)(B-1)(B-2)(C)(D-1) in Table~\ref{tab:vs_baseline} for different sets of models. We observe reasonably degraded performance for 10 and reasonably improved performance for 100 utterances, as expected for the proposed approach (row (D)); the trend for other approaches is similar (row (D) vs rows (A)(B-1)(B-2)(C) for 10, 50, or 100 utterances). This demonstrates the suitability of the proposed approach for personalized acoustic modeling.

    We also attempted integrating the knowledge learned from different token sets with different granularity parameters. This can be easily done by adding more sets of training targets and output networks to the architecture in Fig.~\ref{fig:pt_dnn}. In the joint optimization step during training, we simply use the limited transcribed data to train all the targets, and then use the unlabeled data to train the token state targets. In row (E) of Table~\ref{tab:diff_utt} we chose to integrate token sets $(3,50), (5,200), (13,50), (13,300)$ to capture adequate diversity: slightly better results were obtained than with a single token set (rows (E) vs (D)). We observe that the knowledge learned from different token sets was slightly complementary, but yielded only limited extra improvement, perhaps because they were all extracted in a completely unsupervised way and therefore were less precise. In addition, in comparison to fDLR we see the proposed approach offered more improvements when given fewer transcribed utterances (rows (D) vs (B-1), 2.97\%, 1.24\% and 0.95\% absolute improvements for 10, 50, and 100 utterances respectively). A similar situation is observe with respect to speaker code.

  \section{Conclusions}

  In this paper we propose for personalized acoustic modeling a weakly supervised multi-task deep learning framework based on acoustic tokens discovered from unlabeled data. Output networks for both phoneme states and acoustic token states are jointly learned iteratively during training, such that only very limited amounts of transcribed data need be used with a large set of unlabeled data in the proposed personalized recognizer scenario. Very encouraging initial experimental results were obtained.

  \bibliographystyle{IEEEtran}

  \bibliography{mybib}

\end{document}